\documentclass[
 amsmath,amssymb,
 aps,
]{revtex4-2}

\usepackage{graphicx}
\usepackage{dcolumn}
\usepackage{bm}

\usepackage[margin=1.2in]{geometry}

\begin{document}


\title{Effective potential approach to study hydrodynamics and particle dynamics in Kerr geometry}

\author{Abhrajit Bhattacharjee}
\author{Sandip K. Chakrabarti}%
\affiliation{%
 Indian Centre for Space Physics, 466, Barakhola, Netai Nagar, Kolkata 700099, India
}%


\begin{abstract}
We derive the exact form of effective potential in Kerr geometry from the general relativistic radial momentum equation. The effective potential accurately mimics the general relativistic features, over the entire range of the spin parameter $-1<a<1$. We obtain the exact expression of the rate of dragging of inertial frames that can be used to study the relativistic precession of twisted accretion disks that are formed when the disk outskirts are tilted relative to the equatorial plane of the black hole. We then present an effective potential that provides a simplistic approach to study particle dynamics using physical concepts analogous to the Newtonian physics. We compare the equatorial as well as off-equatorial particle trajectories obtained using our potential with the general relativistic solutions. We find that our approach can capture the salient features of Kerr geometry and is applicable to studies of accretion processes around Kerr black holes.
\end{abstract}

\maketitle


\section{Introduction}
A remarkable consequence of Einstein's theory of general relativity is the  existence of solutions which predicts that enigmatic objects such as black holes should exist. Although they  themselves are the darkest objects in the universe, we find that accreting black holes are  some of the most luminous sources in the sky. Early studies of disk accretion onto black holes were based on the standard disk model \cite{shakura1973black, novikov1973astrophysics} that relied upon a number of simplifying assumptions. In particular, the accreting matter is assumed to have a Keplerian distribution with negligible radial drift velocity and the disk is assumed to terminate at the marginally stable orbit. Moreover, the entire flow is assumed to be subsonic even though black hole accretion flow is necessarily transonic in nature \cite{chakrabarti1990theory}. With the advent of advective solutions it became clear that the standard disk is astrophysically realistic only in a restricted range of its parameters. In particular, it was realized that the flow is not Keplerian everywhere, especially at the inner boundary. The generalized two-component advective flow (TCAF) solution of Chakrabarti \& Titarchuk \cite{chakrabarti1995spectral} that successfully explains the spectral properties of accretion flows incorporate the radiative properties of electron cloud \cite{sunyaev1980comptonization, sunyaev1985comptonization} with the theory of viscous transonic flows \cite{chakrabarti1990theory} which does not stop at the marginally stable orbit but reaches the event horizon of the black hole supersonically after passing through a sonic point.

Relativistic effects play a crucial role in regions close to a black hole. However, to avoid the complexity of general relativity, in the past, a majority of the studies of accretion flows around black holes use the pseudo-Newtonian approach where a simple potential is used to mimic the relativistic effects around black holes through a Newtonian description of hydrodynamics of the accretion and outflows. Paczynsky \& Wiita \cite{paczynsky1980thick} (hereafter PW80) first proposed a pseudo-Newtonian potential that captured the qualitative features of accretion flows around non-rotating black holes quite accurately. Although the potential was chosen in an ad hoc manner, it correctly reproduces the positions of both the marginally stable and the marginally bound circular orbits.

In general, black holes in nature are expected to be rotating \cite{bardeen1970kerr} and physical phenomena around a rotating black hole are usually more difficult to describe than those around a non-rotating black hole. The spin of a black hole causes the relativistic dragging of inertial frames in the vicinity of the black hole, first predicted by Lense \& Thirring \cite{lense1918influence}. Because of this rotation-induced frame-dragging effect, a deviation of particle motion from the equatorial plane results in a precession of the orbital plane of the particle \cite{bardeen1975lense}. Near the event horizon of a rotating black hole, the effects of frame-dragging are enormous, and that it causes all objects to co-rotate with the rotation of the black hole. This should be taken into account in studies of non-equatorial warped disks around rotating black holes. This effect has been experimentally verified in the weak-field limit by satellite experiments in the gravitational field of the spinning Earth \cite{ciufolini2004confirmation, everitt2011gravity}.

As far as the studies of accretion processes around black holes are concerned, especially when the effects of viscosity, magnetic fields and radiative transfer are included, the computations using the full general relativistic framework requires an enormous numerical effort. We wish to extend the pseudo-potential approach of PW80 for Kerr geometry as well, by using reasonably simple choice of effective potentials to study both fluid dynamics and particle dynamics around Kerr black holes. Recently, Bhattacharjee et al. \cite{bhattacharjee2022transonic} has obtained an effective potential from the radial momentum equation in Kerr geometry, valid on and near the equatorial plane of a Kerr black hole and free from any a priori constraints, that accurately reproduces the positions of both the marginally stable and the marginally bound circular orbits for the entire range of the spin parameter $-1<a<1$. Moreover, a study of the transonic properties of inviscid accretion flows using the effective potential revealed that general relativistic solutions could be reproduced very accurately. In this paper, we generalize the effective potential to include the $\theta$-dependence, thereby extending the range of applicability to outside the equatorial plane which appears most convenient for modelling thick disks around rotating black holes. The effective potential is obtained by integrating the radial force in the corotating frame in Kerr geometry and is found to describe hydrodynamics in Kerr geometry very accurately. We can use the effective potential, within the pseudo-Kerr formalism, to study transonic properties of accretion flows around black holes and fit data with the TCAF solution to extract both mass and spin parameters of black hole candidates. We then obtain an exact result regarding the frame-dragging rate in the strong gravity regime that could be used to examine the effects of frame-dragging in theoretical models of accretion disks around black holes. Although the same result was obtained earlier \cite{chakraborty2014strong}, we use a different approach to derive the frame-dragging rate. Moreover, particle dynamics in the vicinity of black holes can give new insights about the structure of the spacetime. Recently, Shakura \& Lipunova \cite{shakura2018logarithmic} has obtained the equations of motion of a test particle in Schwarzschild geometry, using a logarithmic potential, which are identical to those derived using general relativity. We propose an effective potential that incorporate the properties of particle trajectories off the equatorial plane as well and extend their approach to Kerr geometry. We then obtain the equations of motion of a test particle, solve them and study the nature of equatorial and off-equatorial particle trajectories using the effective potential. We find that the characteristics of the particle trajectories are reproduced very accurately. The advantage of using this pseudo-potential approach is that physical concepts analogous to Newtonian physics can be used and the results obtained would be as close to the true general relativistic results as possible.

In Section 2, we derive the generalized effective potential that can be used to study hydrodynamics in Kerr geometry. In Section 3, we derive the exact expression for the frame-dragging rate. In Section 4, we present the effective potential for particle dynamics, solve the equations of motion of test particles using an approach analogous to the Newtonian treatment and compare the results with the corresponding general relativistic solutions. We also briefly study the effects of frame-dragging on the trajectory of a test particle. Finally, in Section 5, we make the concluding remarks.

\section{Effective Potential for Hydrodynamics}
We adopt the geometric system of units $G=M=c=1$, where $G$ is the gravitational constant, $M$ is the mass of the black hole and $c$ is the speed of light. This implies that the units of mass, length and time are $M$, $GM/c^2$ and $GM/c^3$, respectively. The signature of the metric tensor is $(-+++)$. The structure of the spacetime around a rotating black hole is described by the Kerr metric, expressed in the Boyer-Lindquist coordinates $x^\mu=(t,r,\theta,\phi)$ as
 \begin{equation}
     ds^2=-e^{2\nu}dt^2+e^{2\psi}(d\phi-\omega dt)^2+e^{2\lambda}dr^2+e^{2\mu}d\theta^2,
 \end{equation}
 with the functions
 \begin{equation}
    e^{2\nu} = \Sigma\Delta/A,\quad e^{2\psi} = A\sin^2\theta/\Sigma,\quad e^{2\lambda} = \Sigma/\Delta,\quad e^{2\mu} = \Sigma,\quad \omega = 2ar/A.
\end{equation}

Here, $\Delta=r^2-2r+a^2$, $\Sigma=r^2+a^2\cos^2\theta$, $A=(r^2+a^2)^2-a^2\Delta\sin^2\theta$ and $a$ is the spin parameter, the specific angular momentum ($a=J/M$) of the black hole, such that $0\leq|a|\leq1$. The function $\omega=-g_{t\phi}/g_{\phi\phi}$ describes the dragging of inertial frames due to the presence of the off-diagonal term $g_{t\phi}$ in the metric. All physical objects are dragged into orbital motion in the same direction as the rotation of the black hole. The dragging effect is stronger as one moves closer to the black hole. 

The event horizon of the black hole is located at the outer root of the equation $\Delta=0$, i.e., at $r_h=1+(1-a^2)^{1/2}$ for all $\theta,\phi$. We shall restrict ourselves to the region outside the event horizon, with $\Delta>0$. The ergosphere, defined by the condition $g_{tt}=0$, is the surface $r_e = 1+(1-a^2\cos^2\theta)^{1/2}$ that lies outside the event horizon and corresponds to a surface of infinite redshift. It intersects the event horizon only at the poles $\theta=0$ and $\theta=\pi$. For all other latitudes $r_e>r_h$ and the region $r_h<r<r_e$ is called the ergoregion. The contravariant components of the Kerr metric are
\begin{eqnarray}
     g^{tt} = -e^{-2\nu},\quad g^{rr} = e^{-2\lambda},\quad g^{\theta\theta} = e^{-2\mu}, \nonumber \\
     g^{t\phi} = -\omega e^{-2\nu},\quad g^{\phi\phi} = e^{-2\psi}-\omega^2 e^{-2\nu}.
\end{eqnarray}

The Kerr metric is stationary (i.e., time-translation invariant) and axially symmetric. This requires that the metric coefficients $g_{\mu\nu}$ be independent of $t$ and $\phi$ and allows the existence of two Killing vector fields given by
\begin{equation}
    \xi_{(t)}=\left(\frac{\partial}{\partial t}\right)_{(r,\theta,\phi)} \quad \mathrm{and} \quad \xi_{(\phi)}=\left(\frac{\partial}{\partial\phi}\right)_{(t,r,\theta)},
\end{equation}
which are the generators of the corresponding symmetry transformations. Most of the geometrical features of the Kerr spacetime are directly tied to the behavior of these two Killing vector fields.

We consider that the motion of the fluid is characterized by the 4-velocity field
\begin{equation}
    u^\mu = (u^t,u^r,0,u^\phi)
\end{equation}
that satisfies the normalization condition
\begin{equation}
    u_\mu u^\mu = -1,
\end{equation}
where we assume that the $\theta-$component of the 4-velocity is negligible  which is valid for flows in vertical equilibrium.

 The conserved specific angular momentum of the flow is defined as
\begin{equation}
    l = \mathcal{L}/\mathcal{E} = -u_\phi/u_t
\end{equation}
and the angular velocity $\Omega=u^\phi/u^t$ of the flow measured by a stationary observer is then related to $l$ through the relation
\begin{equation}
    \Omega=-\frac{g_{t\phi}+lg_{tt}}{g_{\phi\phi}+lg_{t\phi}}.
\end{equation}
Using Eqs. (6) and (8), we find that the expression for $u^t$ is
\begin{equation}
    u^t = \frac{A^{1/2}\gamma_\phi}{\Delta}\left[\frac{\Delta}{\Sigma}+(u^r)^2\right]^{1/2},
\end{equation}
where 
\begin{equation}
    \gamma_\phi = \left[1-\frac{A^2\sin^2\theta}{\Sigma^2\Delta}(\Omega-\omega)^2\right]^{-1/2}
\end{equation}
is the Lorentz factor associated with azimuthal motion with respect to the locally non-rotating frame (LNRF) that rotates with the coordinate angular velocity $\omega$ and characterized by the world lines $r=const$, $\theta=const$, $\phi=\omega t+const$ \cite{bardeen1972rotating}. Note that, the expression of $u^t$ in Eq. (9) is a function of both $r$ and $\theta$.

Here, we emphasize that the proper time $\tau$ of the LNRF is not equivalent to the coordinate time $t$, and the ratio of their interval is the lapse function $\alpha$ given by
\begin{equation}
    \alpha = \frac{d\tau}{dt} = \left(\frac{\Sigma\Delta}{A}\right)^{1/2}.
\end{equation}

We assume a perfect fluid whose properties in its local rest frame are described by the pressure $p$, the rest mass density $\rho$ and the specific internal energy $\epsilon$, such that the specific enthalpy is $h = 1+\epsilon+p/\rho$. We write the stress-energy tensor of the fluid as
\begin{equation}
    T_{\mu\nu} = h\rho u_\mu u_\nu + pg_{\mu\nu},
\end{equation}
that satisfies the equations of motion
\begin{equation}
    \nabla_\mu T^{\mu\nu}=0,
\end{equation}
where $\nabla_\mu$ is the covariant derivative. The continuity equation can be written as
\begin{equation}
    \nabla_\mu(\rho u^\mu)=0.
\end{equation}
The conserved quantities in a stationary and axially symmetric spacetime are 
\begin{equation}
    \mathcal{E}=hu_t \quad \mathrm{and} \quad \mathcal{L}=-hu_\phi,
\end{equation}
where $\mathcal{E}$ and $\mathcal{L}$ are conserved energy and conserved angular momentum of the flow, respectively, $u_t$ is the specific binding energy and $u_\phi$ is the azimuthal component of the unit flow vector.

Projecting Eq. (13) onto the space orthonormal to $u^\mu$ using the projection tensor, $h_{\mu\nu}=g_{\mu\nu}+u_\mu u_\nu$, yields the relativistic Euler equations
\begin{equation}
    h\rho u^\mu\nabla_\mu u^\nu + (g^{\mu\nu} + u^\mu u^\nu)\partial_\mu p = 0.
\end{equation}

We can write the radial component of $u^\mu\nabla_\mu u^\nu$ as
\begin{equation}
    u^\mu\nabla_\mu u^r = u^r\frac{\partial u^r}{\partial r} + \Gamma^r_{rr}(u^r)^2 + \left[\Omega^2\Gamma^r_{\phi\phi} + 2\Omega\Gamma^r_{t\phi} + \Gamma^r_{tt}\right](u^t)^2
\end{equation}

Substituting the expression of $u^t$ (from Eq. 9) and the connection coefficients in the above equation, we obtain the radial Euler equation which can be written as
\begin{eqnarray}
    u^r\frac{\partial u^r}{\partial r} + \frac{1}{\Sigma\Delta}\left[A\gamma_\phi^2\mathcal{K} + r\Delta -\Sigma(r-1)\right](u^r)^2 \nonumber \\ + \frac{A\gamma_\phi^2\mathcal{K}}{\Sigma^2} + \frac{1}{h\rho}\left[\frac{\Delta}{\Sigma}+(u^r)^2\right]\frac{\partial p}{\partial r} = 0,
\end{eqnarray}
where $\gamma_\phi$ is the Lorentz factor defined in Eq. (10) and 
\begin{equation}
    \mathcal{K} = \frac{r^2-a^2\cos^2\theta}{\Sigma^2}(\Omega a\sin^2\theta -1)^2 -\Omega^2r\sin^2\theta.
\end{equation}

This is a generalization of the radial momentum equation used by Chakrabarti \cite{chakrabarti1996global} to study transonic properties of accretion flows in full general relativity.

In case of motion in the equatorial plane ($\theta=\pi/2$), the expression of $\mathcal{K}$ reduces to the form
\begin{equation}
    \mathcal{K} = \frac{1}{r^2}\left(1-\frac{\Omega}{\Omega_+}\right)\left(1-\frac{\Omega}{\Omega_-}\right),
\end{equation}
where the angular velocities $\Omega_\pm = \pm(r^{3/2}\pm a)^{-1}$ corresponds to prograde ($+$) and retrograde ($-$) Keplerian orbits.

It is convenient to choose the corotating frame (CRF), the frame rotating with the same angular velocity $\Omega$ as the fluid, in which the flow velocity $v$ of the fluid is related to $u^r$ through the relation
\begin{equation}
    \gamma_v v = \frac{v}{(1-v^2)^{1/2}} = u^r g_{rr}^{1/2}.
\end{equation}
and we obtain the expression of $v$ from this equation as
\begin{equation}
    v = \left[1 + \frac{\Delta}{\Sigma(u^r)^2}\right]^{-1/2}.
\end{equation}

\begin{figure}[t]
    \centering
    \includegraphics[scale=0.35]{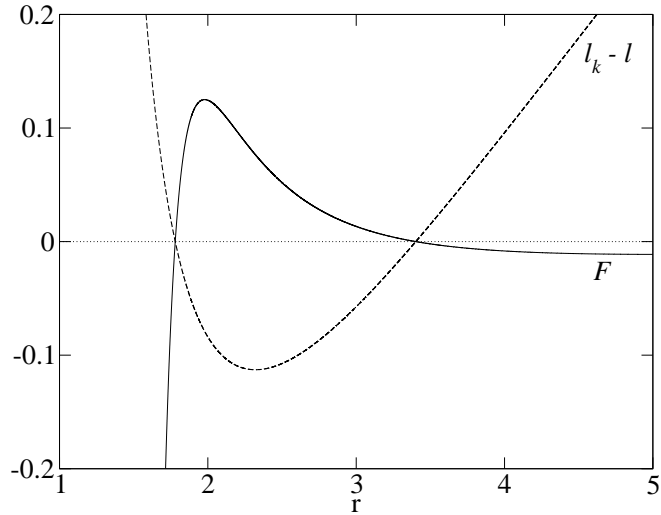}
    \caption{Force (solid curve) and Keplerian distribution of angular momentum (dashed curve) for spin parameter $a=0.9$. The angular momentum $l$ for the force is chosen to be $l=2.6$ and $l_k-l$ curve shows that the force vanishes at the Keplerian orbits where $l=l_k$.}
\end{figure}

Consequently, $v=1$ at the horizon ($\Delta=0$), independent of the mass and spin of the black hole. This means that the infall velocity at the horizon equals the speed of light and the flow is supersonic at the horizon, since the sound speed is always much less than the speed of light. Thus, black hole accretion is necessarily a transonic process \cite{chakrabarti1990theory}.

We use Eq. (22) to rewrite Eq. (18) in the CRF as
\begin{equation}
    \gamma_v^2 v\frac{\partial v}{\partial r} + \frac{A\gamma_\phi^2\mathcal{K}}{\Sigma\Delta} + \frac{1}{h\rho}\frac{\partial p}{\partial r} = 0.
\end{equation}

The second term of this equation, in analogy with Newtonian hydrodynamics, can be identified as the gradient of an effective potential. Using this, the radial force can be written as
\begin{equation}
    F(r) = -\frac{\partial\Phi_\mathrm{eff}}{\partial r} = -\frac{A\gamma_\phi^2\mathcal{K}}{\Sigma\Delta}
\end{equation}
and the effective potential is found to be
\begin{equation}
    \Phi_\mathrm{eff} = 1 + \frac{1}{2}\ln\left[\frac{\Delta\Sigma\sin^2\theta}{(A+a^2l^2-4alr)\sin^2\theta-l^2\Delta}\right].
\end{equation}

This is the most general form of the effective potential in Kerr geometry. The corresponding gravitational potential obtained from $\Phi_\mathrm{eff}$ by using $l=0$ in the above expression and excluding the rest-mass energy is
\begin{equation}
    \Phi_\mathrm{g} = \frac{1}{2}\ln\left(\frac{\Delta\Sigma}{A}\right).
\end{equation}
This is the exact relativistic gravitational potential that can be expressed in terms of the lapse function $\alpha$, such that the gravitational force on a test particle measured in the LNRF is $\mathbf{F}=-\nabla\ln\alpha$ \cite{thorne1986black}. At the equatorial plane ($\theta=\pi/2$), $\Phi_\mathrm{eff}$ takes the form \cite{bhattacharjee2022transonic}
\begin{equation}
    \Phi_\mathrm{eff} = 1 + \frac{1}{2}\ln\left(\frac{r\Delta}{r^3+(a^2-l^2)r+2(a-l)^2}\right).
\end{equation}
This potential is found to reproduce correctly the locations of both the marginally stable and marginally bound orbits for the entire range of the spin parameter in Kerr geometry. So, the effective potential (Eq. 25) can be used to study equatorial as well as off-equatorial motion of matter around rotating black holes very accurately.

An important aspect of motion of matter around a black hole is that the radial force reverses sign several times when the angular momentum of the matter is larger than the marginally stable value corresponding to the minimum of the Keplerian distribution of angular momentum. This implies the existence of more than one Keplerian orbit, in contrast to the Newtonian case where the force reverses sign only once \cite{chakrabarti1993reversal}. In Fig. 1, we plot the force $F$ from Eq. (24) for $a=0.9,\, l=2.6$ and the Keplerian distribution of angular momentum. It is found that the force vanishes at two locations for a given angular momentum corresponding to two Keplerian orbits $l=l_k$, as expected. This means that the force is repulsive in the region between the two Keplerian orbits and attractive elsewhere. Since the  Keplerian orbits are associated with the extrema of the effective potential (Eq. 25) where the radial force vanishes, this reversal of the force can have important applications in studies of accretion disks around black holes.

\section{Frame Dragging}
The dragging of inertial frames, or frame-dragging, is a natural consequence of spacetime curvature in the vicinity of a rotating black hole that causes the induced rotation of locally inertial frames with respect to the inertial frame of a distant observer whose axes are aligned with the ``fixed stars" at spatial infinity. This is a general relativistic feature that occurs in all stationary spacetimes with $g_{t\phi}\neq0$ and has no analog in Newtonian gravitational theory.

From the point of view of a distant observer, a freely falling test particle with zero angular momentum would co-rotate with the geometry and, as a consequence, would acquire a non-vanishing angular velocity. Frame-dragging has an intriguing influence on the off-equatorial motion of test particles, the Lense-Thirring effect, that causes the gyroscopic precession of the orbital plane around the spin axis of a Kerr black hole \cite{lense1918influence, bardeen1975lense}. 

We choose a family of static observers having 4-velocity everywhere parallel to the time-like Killing vector field $\partial_t$:
\begin{equation}
    u^\alpha=(-g_{tt})^{-1/2}\partial_t^\alpha.
\end{equation}
Such observers are kept fixed relative to distant observers at spatial infinity and are time-like only outside the ergoregion where $g_{tt}<0$. Thus, the static observers are characterized by constant values of ($r,\theta,\phi$) and they are located at fixed spatial coordinates.

It is convenient, however, to cast the spacetime metric into the form \cite{landau2013classical}
\begin{equation}
    ds^2 = g_{tt}(dt-g_idx^i)^2 + \gamma_{ij}dx^idx^j,
\end{equation}
with $i,j=1,2,3$, where $g_i=-g_{ti}/g_{tt}$ are the components of a 3-dimensional vector $\mathbf{g}$ and 
\begin{equation}
    \gamma_{ij}=\left(g_{ij}-\frac{g_{ti}g_{tj}}{g_{tt}}\right)
\end{equation}
is the spatial metric of a 3-space, on which the line element $dl^2=\gamma_{ij}dx^idx^j$ gives the infinitesimal spatial distance between any two nearby events. The 3-dimensional metric $\gamma_{ij}$ determines the spatial geometry and can be used to calculate the norm of 3-vectors and projections between them. Accordingly, it is convenient to choose the set of basis vectors $\{\mathbf{e_i}\}$ associated with the static observers to be \cite{thorne1971relativistic}
\begin{equation}
    \mathbf{e_1} = \frac{\partial}{\partial r},\quad
    \mathbf{e_2} = \frac{\partial}{\partial\theta},\quad 
    \mathbf{e_3} = \frac{\partial}{\partial\phi} - \left(\frac{g_{t\phi}}{g_{tt}}\right)\frac{\partial}{\partial t}.
\end{equation}

We now consider a locally inertial frame that is momentarily at rest relative to a static observer. The direct manifestation of frame-dragging is then the rotation of the space axes of the locally inertial frame relative to the basis vectors $\{\mathbf{e_i}\}$ of the static observer. Since the frame-dragging rate varies with position, we have a differential rotation between adjacent frames and this rate of differential rotation is

\begin{equation}
    \mathbf{\Omega} = -\frac{1}{2}(-g_{tt})^{1/2}(\mathbf{\nabla}\times\mathbf{g}).
\end{equation}

This equation is written in the 3-vector notation in the frame of the static observers. Thus, the frame-dragging rate is obtained as
\begin{equation}
    \mathbf{\Omega} = \frac{a}{\Sigma^2(\Sigma-2r)}\left[2r\Delta\cos\theta\frac{\partial}{\partial r} - (\Sigma-2r^2)\sin\theta\frac{\partial}{\partial\theta}\right].
\end{equation}

The norm of this vector is found to be
\begin{equation}
    \Omega = \frac{a\left[4r^2\Delta\cos^2\theta+(\Sigma-2r^2)^2\sin^2\theta\right]^{1/2}}{\Sigma^{3/2}(\Sigma-2r)} .
\end{equation}

\begin{figure}[t]
    \centering
    \includegraphics[scale=0.35]{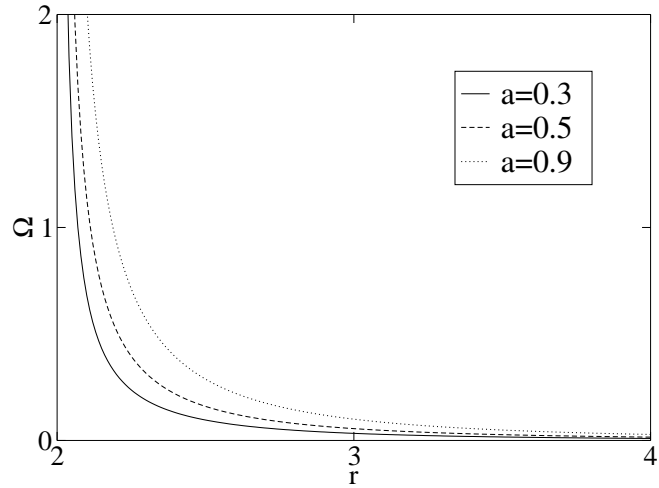}
    \caption{Variation of the frame-dragging rate $\Omega$ (Eq. 34) as a function of the radial coordinate at the equatorial plane ($\theta=\pi/2$) of a Kerr black hole for values of the spin parameter $a=0.3,0.5,0.9$.}
\end{figure}

This is the exact frame-dragging rate where no weak-field approximation has been made. This expression is finite and regular outside the ergoregion of a Kerr black hole. Although the frame-dragging rate has been derived earlier \cite{chakraborty2014strong}, we use a different approach to arrive at the results. In Fig. 2, we present the variation of $\Omega$ with the distance from the black hole for $a=0.3,0.5,0.9$. Clearly, the frame-dragging rate gradually increases as one moves closer to the black hole, and the effect is stronger for a greater value of the spin parameter. 

Frame-dragging is usually manifested in terms of its local effects on gyroscopes. A test gyroscope with spin angular momentum $\mathbf{S}$ appears to precess with an angular velocity $\mathbf{\Omega}$ relative to an asymptotic inertial frame, and the rate of change of $\mathbf{S}$ is $d\mathbf{S}/dt=\mathbf{\Omega}\times\mathbf{S}$.

In the limit of large $r$, i.e., in the weak-field limit, Eq. (33) reduces to
\begin{equation}
    \mathbf{\Omega} = \frac{a}{r^3}\left[2\cos\theta\frac{\partial}{\partial r} + \frac{\sin\theta}{r}\frac{\partial}{\partial\theta}\right],
\end{equation}
whose norm is $\Omega=(a/r^3)\sqrt{1+3\cos^2\theta}$. In the usual 3-vector notation, Eq. (35) can be expressed as
\begin{equation}
    \mathbf{\Omega} = \frac{1}{r^3}\left[-\mathbf{J}+\frac{3(\mathbf{J}\cdot\mathbf{r})\mathbf{r}}{r^2}\right],
\end{equation}
where $\mathbf{J}$ is the angular momentum of the black hole and $\mathbf{r}$ is the position vector of the gyroscope. This is the standard result in the weak-field approximation.\cite{lense1918influence} 

Notice that near the polar region the precession is in the same direction as the black hole, but in the opposite direction near the equatorial plane. For small velocities of motion ($|\mathbf{v}|\ll 1$), it appears that a force analogous to the Coriolis force, acts on a test particle in the frame of the static observers and the acceleration is given by \cite{landau2013classical}
\begin{equation}
    \mathbf{a} = -\nabla(\ln\sqrt{-g_{tt}})+2\mathbf{\Omega}\times\mathbf{v}.
\end{equation}
Here, the first term is simply the gravitational acceleration, and the second term is the Coriolis acceleration that is caused by the dragging of the inertial frames relative to the static observers.

\section{Particle Dynamics}
We now investigate the nature of equatorial as well as off-equatorial particle trajectories which is of some practical interest in studies of accretion disks around black holes. It is convenient to introduce a set of local observers having a finite angular velocity $\omega(r,\theta)=d\phi/dt$ relative to static observers at infinity \cite{bardeen1972rotating}. Such observers have no angular momentum with respect to the black hole and they observe an unchanging geometry of the spacetime. The spacetime is locally flat in the frame of such local observers and can be described by the Minkowski metric.

From the metric (Eq. 1), we find that the distance element $d\tilde{r}$ along the radial direction, measured by a local observer, is longer relative to the corresponding coordinate element $dr$:
\begin{equation}
    d\tilde{r} = \left(\frac{\Sigma}{\Delta}\right)^{1/2}dr.
\end{equation}

Let us consider a test particle moving with velocity $\vec{v}$, as measured by the local observers. Then, the time measured by a local observer $\tau$ is related to the time measured in the frame of the moving particle $\tau_p$ as
\begin{equation}
    \frac{d\tau_p}{d\tau} = (1-v^2)^{1/2},
\end{equation}
where $v^2=\delta_{(i)(j)}v^{(i)}v^{(j)}; i,j=r,\theta,\phi$, in terms of the physical (not the coordinate) components of the velocity in the local frame.

We can write the general expression for the squared 4-momentum of a test particle as
\begin{equation}
    p^\mu p_\mu = g^{\mu\nu}p_\mu p_\nu = -1,
\end{equation}
which explicitly stands for the expression
\begin{equation}
    -\frac{A(p_t+\omega p_\phi)^2}{\Sigma\Delta} + \frac{\Sigma p_\phi^2}{A\sin^2\theta} + \frac{\Delta p_r^2}{\Sigma} + \frac{p_\theta^2}{\Sigma} = -1.
\end{equation}

In the general case, the energy $E=-p_t$ of a particle and the projection of its angular momentum $l=p_\phi$ along the symmetry axis of the black hole are conserved quantities. The physical values of the energy and the three spatial components of momentum of the particle as measured by the local observers can be written as \cite{shakura1987geodesics}
\begin{equation}
    p^{(t)}=\gamma \quad \mathrm{and} \quad p^{(i)}=\gamma v^{(i)},
\end{equation}
where $\gamma=(1-v^2)^{-1/2}$ and $i=r,\theta,\phi$. Since the spacetime appears flat for local observers, Eq. (40) can be expressed as
\begin{equation}
    p^{(\alpha)}p_{(\alpha)}=\eta_{(\alpha)(\beta)}p^{(\alpha)}p^{(\beta)}=-1,
\end{equation}
where $\eta_{(\alpha)(\beta)}=diag(-1,1,1,1)$. Using Eqs. (41)-(43), we find that the expression for the Lorentz factor is
\begin{equation}
    \gamma = \sqrt{\frac{A}{\Sigma\Delta}}(E-\omega l)
\end{equation}
and the physical velocity components that can be used by local observers to understand the general properties of test particle motion in their frame are
\begin{eqnarray}
    v^{(r)} &=& \sqrt{\frac{\Sigma}{\Delta}}\left(\frac{dr}{d\tau}\right), \\
    v^{(\theta)} &=& \sqrt{\Sigma}\left(\frac{d\theta}{d\tau}\right), \\
    v^{(\phi)} &=& \frac{\Sigma l\sqrt{\Delta}}{(E-\omega l)A\sin\theta}.
\end{eqnarray}

It is important to note that physical measurements carried out by the local observers pays no attention to the coordinate time $t$, rather they use their proper time $\tau$, related through Eq. (39), to the particle's proper time $\tau_p$. We can rewrite Eq. (44) in the form
\begin{equation}
    E = \alpha E_\mathrm{loc} + \omega l,
\end{equation}
where $\alpha$ is the lapse function (Eq. 11) and $E_\mathrm{loc}=\gamma$ is the locally measured energy of the particle. The term $\omega l$ in the above equation is due to the orbital motion of the local observers relative to distant observers and $E$ defines the notion of energy-at-infinity which is conserved along the trajectory of a particle, while $E_\mathrm{loc}$ varies in the gravitational field of the black hole \cite{thorne1986black}. In case of a non-rotating black hole, we find that Eq. (47) reduces to the form
\begin{equation}
    l=\frac{rv^{(\phi)}}{\sqrt{1-v^2}}
\end{equation}
at the equatorial plane. This is the usual expression for the conserved angular momentum of a test particle in a spherically-symmetric gravitational potential in special relativity. Now, taking into consideration that the velocity $\vec{v}$ of the particle satisfies $v^2=(v^{(r)})^2+(v^{(\theta)})^2+(v^{(\phi)})^2$, Eqs. (44)-(47), with the aid of Eq. (39) gives
\begin{equation}
    \frac{\Sigma}{\Delta}\left(\frac{dr}{d\tau_p}\right)^2 + \Sigma\left(\frac{d\theta}{d\tau_p}\right)^2 = \frac{A}{\Sigma\Delta}(E-\omega l)^2 - \frac{\Sigma l^2}{A\sin^2\theta} - 1.
\end{equation}

This equation corresponds to the exact solution of the motion of a test particle with energy $E$ in general relativity \cite{chandrasekhar1983mathematical}. Although simplification of this equation reveals that there are terms which resemble the contributions from spin-orbit coupling (proportional to $al$), as well as gravitational (terms without $l$) and centrifugal (proportional to $l^2$) energies, it is not possible to directly interpret them as such in general relativity. However, we can use this to propose an approximate Newtonian effective potential that could mimic the relativistic effects around black holes. Using the same spirit of `ad hoc'-ness as our earlier attempts \cite{chakrabarti1992newtonian, chakrabarti2006studies}, we propose an effective potential of the form:

\begin{equation}
    V_\mathrm{eff} = 1 - \frac{r}{\Delta}\left(1+\frac{a^2}{r^2}\right) + \frac{2al}{r\Delta} + \frac{l^2}{2r^2\sin^2\theta}\left(1-\frac{a^2\sin^2\theta}{\Delta}\right).
\end{equation}

It is easy to understand the significance of the individual terms. Here, the first term is clearly the rest mass of the particle. The second term is the gravitational potential energy which is singular at the event horizon $r_h=1+\sqrt{1-a^2}$. This is not a serious issue in regions of interest $r\ge r_{ms}>r_h$ for accretion flows around black holes. The third term corresponds to the coupling between the spin $a$ of the black hole and the specific angular momentum $l$ of the particle that mimics the spin-orbit coupling, whereas the fourth term is the centrifugal potential that describes the rotational energy of the particle. Note that, in the Schwarzschild limit, Eq. (51) takes the form $V_\mathrm{eff}=1-1/(r-2)+l^2/(2r^2\sin^2\theta)$, where the second term is the well known PW80 potential. Thus, the motion in $r$ and $\theta$ can be thought of as motion in the effective potential $V_\mathrm{eff}$, whose form is chosen in such a way that it simplifies to the PW80 form for $a=0$ and which, at the same time, reproduces the nature of frame-dragging at the horizon.

\begin{figure}[t]
    \centering 
    \includegraphics[scale=0.45]{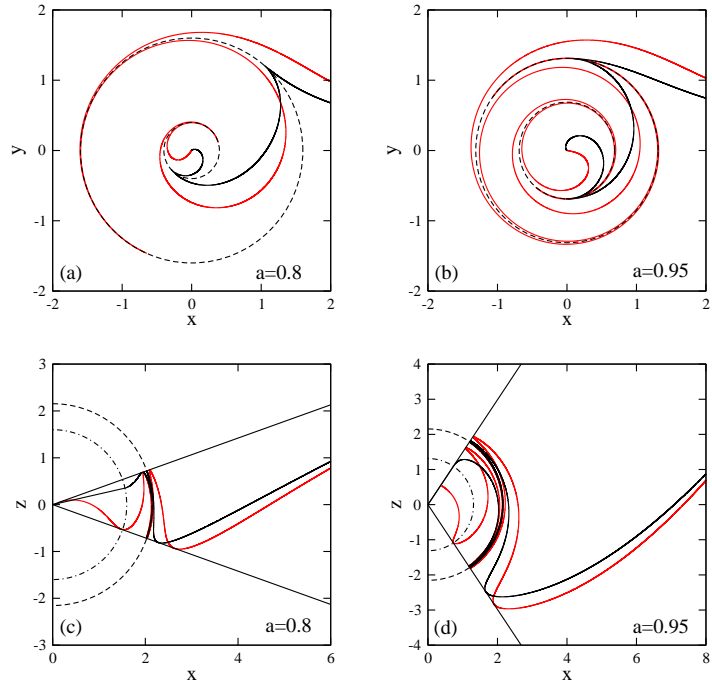}
    \caption{(a-b) Trajectories of particles with zero angular momentum in the equatorial plane of a Kerr black hole with spin parameter (a) $a=0.8$ and (b) $a=0.95$ plotted in the ($x,y$)-plane. The locations of the horizons are shown by the dashed circles. (c-d) Projections of off-equatorial particle trajectories for the parameters (c) $a=0.8,l=2.79$ and (d) $a=0.95,l=1.76$ plotted in the ($x,z$)-plane. The dashed circle indicates the radial coordinate $r_s=2.154$ where the radial velocity is zero, chosen same in both the cases. The motion is bounded within the region defined by $\theta_s$ and $\pi-\theta_s$ and the values of $\theta_s$ are (c) $\theta_s=1.23$ rad and (d) $\theta_s=0.59$ rad. The comparison with the exact general relativistic solutions (the red curves) are illustrated.}
\end{figure}

\begin{figure}[t]
    \centering
    \includegraphics[scale=0.3]{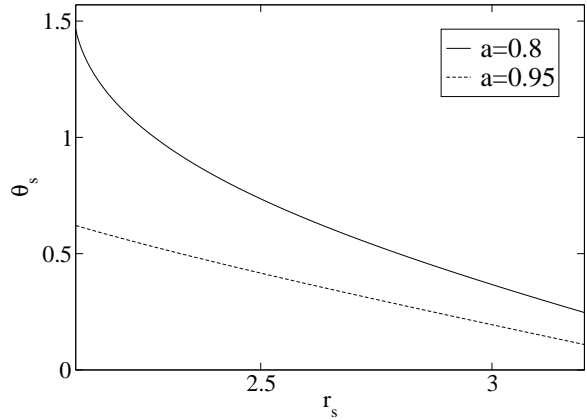}
    \caption{The variation of $\theta_s$ with the radial coordinate $r_s$ for values of the spin parameter $a=0.8,0.95$ for direct orbits.}
\end{figure}

Here, we wish to stress that, separating the relativistic force into distinct components is not feasible in a fully general relativistic framework \cite{chakrabarti1993reversal} and as a result, discussing the individual components of energy is meaningless. However, using Newtonian physics, this potential can be used in approximating relativistic effects and describing quite general test-particle motion, including frame dragging effects, around Kerr black holes.

To study the nature of particle trajectories, we can write the Hamiltonian in spherical coordinates as
\begin{equation}
    \mathcal{H} = \frac{1}{2}\left(p_r^2+\frac{p_\theta^2}{r^2}\right) + V_\mathrm{eff},
\end{equation}
where $p_r=\dot{r}$ and $p_\theta=r^2\dot{\theta}$. The conjugate momentum $p_\phi$ associated with the cyclic coordinate $\phi$ is the conserved angular momentum $l=p_\phi=r^2\sin^2\theta\dot{\phi}$ of the particle. Since the Hamiltonian $\mathcal{H}$ does not depend explicitly on time, $\mathcal{H}$ is physically the conserved energy $E$ of the particle. We consider here, for simplicity, that a test particle starts at rest from infinity, in which case $E=1$. The $r$ and $\theta$ components of the equation of motion is obtained by the method of separation of variables as 
\begin{equation}
    \dot{r} = \left[\frac{1}{\Delta}\left\{2r\left(1+\frac{a^2}{r^2}\right)+\frac{a^2l^2}{r^2}-\frac{4al}{r}\right\}-\frac{\alpha_\theta}{r^2}\right]^{1/2} \equiv f(r)
\end{equation}
and 
\begin{equation}
    \dot{\theta} = \frac{1}{r^2}\left[\alpha_\theta-\frac{l^2}{\sin^2\theta}\right]^{1/2} \equiv \frac{1}{r^2}g(\theta),
\end{equation}
where $\alpha_\theta$ is the separation constant that can be expressed as $\alpha_\theta=p_\theta^2+l^2\csc^2\theta$. This is equivalent to the so-called Carter's constant in our Newtonian approach that incorporates the contribution from motion in the $\theta-$direction. When the motion of the particle is constrained to the equatorial plane (for which $\dot{\theta}=0$ and $\theta=\pi/2$), the equation of motion for the azimuthal coordinate $\phi$ is given by
\begin{equation}
    \dot{\phi} = \frac{1}{\Delta}\left[\frac{2a}{r}+\left(1-\frac{2}{r}\right)l\right].
\end{equation}

We solve Eqs. (53)-(55) using the fourth order Runge-Kutta method  and consider only the special cases for which the radial coordinate remains constant. We illustrate, in Fig. 3, examples of solutions obtained using our approach and comparison with the corresponding general relativistic solutions \cite{chandrasekhar1983mathematical}. In Fig. 3(a-b), we plot the trajectories of particles having zero angular momentum, i.e. $l=0$, at the equatorial plane of a Kerr black hole for (a) $a=0.8$ and (b) $a=0.95$. The axes are $x=r\cos\phi$ and $y=r\sin\phi$. The dashed circles indicates the locations of the two horizons. Since the angular momentum of the particle is chosen to be zero, the particle is initially radially ingoing at infinity. However, it gradually picks up angular velocity due to frame-dragging, although the conserved angular momentum of the particle remains zero. The effect of frame-dragging and its spin-dependence is clearly illustrated in the trajectory of the particle. In Fig. 3(c-d), we plot the projections of the off-equatorial particle trajectories in the meridional plane for (c) $a=0.8,\, l=2.79$ and (d) $a=0.95,\, l=1.76$. The axes are $x=r\sin\theta$ and $z=r\cos\theta$. The particles are released away from the equatorial plane. In both the cases, we assume that the radial velocity of the flow is zero at $r_s=2.154$. This puts a constraint on the constants of motion through the conditions $f(r)=f'(r)=0$. We find that the motion in confined in latitute for non-zero $l$ and the values of $\theta_s$ (obtained using the condition $g(\theta)=0$) where $\dot{\theta}=0$ are (c) $\theta_s=1.23$ rad and (d) $\theta_s=0.59$ rad, respectively. Thus, the motion is confined within the region defined by the angular coordinates $\theta_s$ and $\pi-\theta_s$. We observe significant deviation only very close to the event horizon for off-equatorial trajectories where the particle rapidly falls into the black hole while crossing the event horizon. However, for astrophysically relevant accretion flows, we are interested only in the region $r>r_h$ and when $l$ is close to the marginally stable and the marginally bound values. In such cases, we find that our results match very well with the exact general relativistic solutions. Note that the deviation from the general relativistic trajectories arises because the potential (Eq. 51) is an approximate Newtonian potential that incorporates relativistic effects, whereas the exact trajectories include non-linear coupling terms whose effects cannot be isolated within a Newtonian framework. In Fig. 4, the variation of $\theta_s$ with $r_s$ is shown for direct orbits for two values of the spin parameter $a=0.8,\, 0.95$. This angle $\theta_s$ determines, at large distances, the orientation of the accretion disk relative to the equatorial plane of a black hole.

\begin{figure}[t]
    \centering
    \includegraphics[scale=0.45]{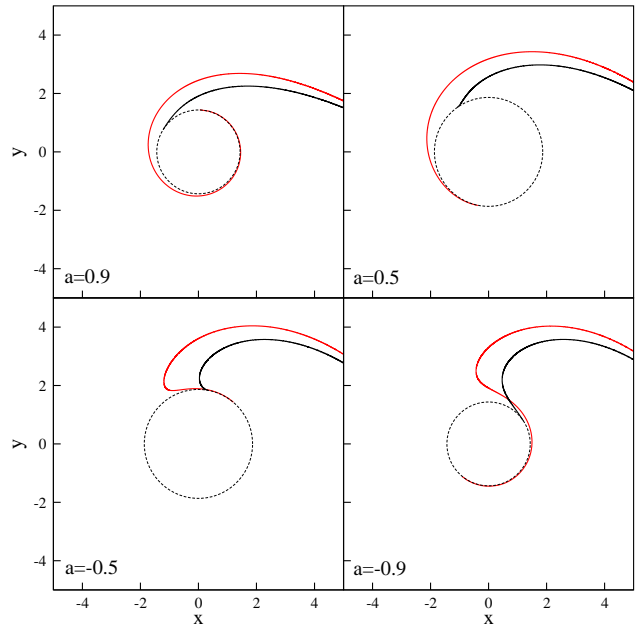}
    \caption{Plots of equatorial particle trajectories with non-zero angular momentum in the region outside the event horizon of a Kerr black hole with values of the spin parameter $a=0.9,0.5,-0.5,-0.9$. The red curves correspond to the exact general relativistic trajectories. The location of the event horizon is shown with a dashed circle.}
\end{figure}

In Fig. 5, we plot the equatorial orbits (both direct and retrograde) outside the event horizon of a Kerr black hole with spin parameter $|a|=0.5,0.9$ for a particle with non-zero angular momentum. The axes are $x=r\cos\phi$ and $y=r\sin\phi$. The dashed circle at the centre of each panel indicates the location of the event horizon of the black hole. It is of particular interest that for retrograde motion ($a<0$),  i.e., when the sign of $l$ is opposite to that of $a$, the angular velocity can flip sign near the horizon. This can be clearly observed in the figure and this means that there is a transition from retrograde to direct motion at some radial coordinate close to the event horizon of the black hole. We find that this reversal of angular velocity for retrograde orbits is also reproduced using our potential which is an important phenomenon leading to observable signatures in studies of accretion disks around rotating black holes.

\section{Concluding Remarks}
In the present paper we have derived the exact effective potential for hydrodynamics in Kerr geometry by identifying the radial force term in the general relativistic radial momentum equation written in the corotating frame. This potential reproduces accurately the locations of both the marginally stable and marginally bound orbits for the entire range of the spin parameter $-1<a<1$. The effective potential for particle dynamics is found to reproduce the nature of equatorial as well as off-equatorial particle trajectories very accurately outside the horizon. Precise measurements of the effects of frame-dragging is crucial because it is expected to cause non-trivial changes in the structure of an accretion disk close to a Kerr black hole, in case the disk is not in the equatorial plane of the black hole, particularly in studies of thick accretion disks \cite{chakrabarti1985natural}. The nature of frame-dragging is also correctly exhibited at the horizon using our potential within the framework of Newtonian physics. We have also presented an exact treatment of frame-dragging valid in the strong gravity regime. Since the accretion disks usually extend to the vicinity of the horizon of a black hole where the effects of gravity is enormous, it will be interesting to investigate the effects of frame-dragging in modelling of accretion disks. Thus, for all practical purposes, our pseudo-Kerr approach captures the salient features of the spacetime around a Kerr black hole with reasonable accuracy. Moreover, we can describe accretion processes around a Kerr black hole within a Newtonian framework, still capturing the strong gravity effects in the vicinity of the black hole. This approach is particularly useful to address complex situations such as problems involving magnetic fields and radiative transfer as has been done using the PW80 potential in the context of non-rotating black holes. Our approach can also be utilized to generalize the models of thick accretion disk \cite{paczynsky1980thick} and to construct shock solutions \cite{chakrabarti1989a, chakrabarti1989b, chakrabarti1995spectral} in accretion flows around Kerr black holes.

Furthermore, time-dependent numerical simulations of accretion flows can be carried out quite easily using Newtonian numerical codes written for flat geometry \cite{ryu1995stable, hawley2001global, giri2013hydrodynamic}. Recently, numerical simulations to investigate the effect of magnetic field on shock waves in magnetized accretion flows around non-rotating black holes modelled using the PW80 potential have been carried out \cite{garain2020effects}. Using our effective potential approach, such studies can be easily extended to Kerr black holes as well. It is well-known that shocks are formed in low angular momentum accretion flows around black holes \cite{chakrabarti1989a, chakrabarti1990theory, chakrabarti1996grand}. Moreover, the resonance oscillation of the shocks are considered to be the driving force behind the so-called quasi-periodic oscillations (QPOs) of the hard X-rays from black hole candidates \cite{chakrabarti2015resonance} and for rapidly rotating black holes, shocks form much closer to the black holes which implies that the QPOs would have higher frequencies. Thus, our pseudo-Kerr formalism can be used to fit data with the TCAF solution that may aid to measure the spin parameter from the observed QPO frequencies of black hole candidates.

\newpage

\begin{acknowledgments}
The authors thank the anonymous reviewer for valuable comments and suggestions that improved the presentation of the paper. The authors acknowledge a grant of the ISRO sponsored RESPOND project (ISRO/RES/2/418/18-19). A.B. also acknowledges a grant towards a Junior Research Scientist position at ICSP from the Govt. of West Bengal, India.
\end{acknowledgments}

\bibliography{paper2}

\end{document}